\documentclass{article}
\begin{document}
\author{Bjoern S. Schmekel \\ Theoretical Astrophysics Center \\University of California, Berkeley, California 94720}
\title{Quasi-local definitions of energy in general relativity}

\maketitle

Defining energy is a surprisingly difficult problem in general relativity. For instance, the energy density of the gravitational field of a planet at a particular point could be determined by a comoving observer measuring the kinetic energy of a freely falling object. Due to the equivalence  
principle both the object and the observer fall at equal rates. Therefore, the observer would
not assign any energy to the object. Other observers like an observer who is at rest with
respect to the planet would measure different values. This raises the question of how energy 
depends on the choice of an observer which violates the philosophy of general relativity
whose tensorial equations are independent of the used reference system. 

In classical electrodynamics the stress-energy tensor is a measure of the energy 
and momentum transported
by the electromagnetic field due to a source distribution $j^\mu$. A similar
construction in general relativity leads to the so-called Bel-Robinson tensor 
$T_{\mu \nu \rho \sigma}$ \cite{Bel:1958,Robinson:1958,Bel:1959} which can be thought of as being
induced by a stress-energy tensor $T_{\mu \nu}$.
Its physical meaning however remains unknown since it does not even have
units of energy density. This is a consequence of the equivalence
principle which equates the gravitational mass (the "charge" of gravity)
with the inertial mass. The source term, i.e. $j_\mu$ in electrodynamics
and $T_{\mu \nu}$ in general relativity, does not contain the energy of the
gravitational field. However, since the equations of general relativity are non-linear
there may be a non-linear contribution to the stress-energy.
For instance, gravitational waves do not pass through each other without distortion.

Due to the absence of Stokes theorem for second ranked tensors conserved quantities
do not exist.

Landau and Lifshitz were able to prove that the stress-energy-momentum pseudotensor
\begin{eqnarray}
16 \pi G t^{\mu \nu} = (-g)^{-1} \left [ (-g) \left ( g^{\mu \nu} g^{\rho \sigma}
- g^{\mu \rho} g^{\nu \sigma} \right ) \right ]_{,\rho \sigma} - 2 G^{\mu \nu}
\end{eqnarray}
is the only symmetric pseudotensor constructed only from the metric
such that the four-divergence of the total stress energy vanishes like
$\left [ (-g)(T^{\mu \nu} + t^{\mu \nu} ) \right ]_{,\mu}=0$
and which also vanishes locally in an inertial frame.
The latter requirement is dictated by the equivalence principle as was mentioned
above. However, $t^{\mu \nu}$ still does not transform as a tensor.

Because of the problems associated with defining a local energy density
it may be easier to make sense of the energy enclosed by a boundary. For regions
of finite extend we expect non-zero values because in general a coordinate transformation
can make the connection coefficients vanish at only one point.

Therefore, it seems the only sensible way to define energy is by defining energy
itself and not energy density. Of course this may seem ugly because a local covariant
and tensorial formulation depends on densities evaluated at a point and its
infinitesimally small neighborhood (in order to compute derivatives). A point
remains a point under a Lorentz transformation, but needless to say the size
of a finite region depends on the observer, so obviously such an energy
will depend on the chosen coordinate system. It is therefore maybe not surprising
that the first useful notions of energy were defined at infinity, i.e. they
enclosed the whole system (cf. ADM mass \cite{ADM:1962}, Bondi mass \cite{Bondi1964}).
Like a point an infinitely large box does not change its size
under a change of observer. 

A successful definition of quasi-local energy (QLE) was given by Brown and York \cite{Brown:1992br}.
A spacelike three-dimensional hypersurface $\Sigma$ is embedded in a four-dimensional spacetime $M$
which satisfies the Einstein field equations. This embedding defines the
"time-direction". Finally, a two-dimensional boundary $B$, which encloses the energy
of the region we are interested in, is embedded in the three-dimensional hypersurface $\Sigma$. 
$\Sigma$ is enclosed by a three-boundary $^3 B$, and their normals are constrained to
be perpendicular to each other. This restriction ensures the time evolution of the system
is consistent with the presence of the fixed boundary $B$.
In classical mechanics the Hamiltonian is given by the variation of the classical action with
respect to the endpoints times minus one. The Brown-York QLE is derived
by considering the change of the classical action under a displacement of the initial
and final hypersurfaces and is given by
\begin{eqnarray}
E = \frac{1}{\kappa} \int_B d^2 x \sqrt{\sigma} \left ( k - k_0 \right )
\end{eqnarray}
where $k$ is the trace of the extrinsic curvature of $B$ and $\sigma$ is the metric of $B$.
The surface gravity is denoted by $\kappa$, and $k_0$ is a reference term 
which sets the energy of flat space to zero. The subtraction term has
been criticized. However, it should be mentioned that the ADM energy makes
reference to flat space as well by using ordinary non-covariant derivatives.

For a Schwarzschild black hole the action can be expressed in terms of the
QLE as follows
\begin{eqnarray}
S = \frac{8\pi}{\kappa} \int (N dt) (r f)
\end{eqnarray}
where $-r f$ is the unreferenced QLE. The metric has been
expressed as $ds^2 = - N^2(r) dt^2 + f^{-2}(r) + r^2 d \Omega^2$
where $N(r)=f(r) = \sqrt{1- 2m/r}$. Because the geodesics of infalling 
objects can be determined both inside and outside of a black hole it 
should be possible to assign a value to the energy of the gravitational 
field in both regions. The definition of $E$ can be continued into 
the region inside the horizon \cite{Lundgren:2006fu}. Since both $N$ and $f$ become imaginary
inside the horizon $-rf$ needs to be multiplied by $i$ in order to
become real. 

Whether the quantity defined above is useful depends on its properties and
whether applications exist. At infinity the QLE is equal to the ADM energy.
Furthermore, it reduces to the Newtonian binding energy in the non-relativistic
limit. In the thermodynamics of black holes the QLE is just the total energy.
Blau and Rollier have shown that the extended Brown-York energy describes
the effective potential of a particle falling into a black hole \cite{Blau:2007wj}.
Also worth mentioning is the small sphere limit \cite{Brown:1998bt}. In this limit
the QLE reduces to the energy of the enclosed matter. The gravitational
binding energy only shows up in higher orders of the radius which emphasizes
the fact that one cannot make sense of the energy of the gravitational field locally.
This may be seen as a hint that point particles do not exist \cite{Lundgren:2006fu}.

The most serious drawback is that fact that not all physically interesting
boundaries $B$ can be embedded in a reference space which is typically
taken to be $\mathcal{R}^3$ leading to a non-existence of the reference term. 
An important example is the horizon of a Kerr black hole. 
Usually only energy differences are important, so the absence of the
reference term might not lead to problems. However, the stability of
flat space rests on the fact that every non-flat spacetime contains more
energy that the flat ground state \cite{Schon:1981vd}. Therefore, Epp \cite{Epp:2000zr}
and subsequently Liu und Yau \cite{Liu:2003bx} considered a modification of
the Brown-York energy which does not need the three-boundary
$^3 B$. Rather, the two-boundary $B$ is embedded directly
into the four-dimensional spacetime $M$, and the orthogonality condition
is not applicable anymore. Such an embedding does always exist.
However, it is not unique. The absence of the orthogonality condition
implies that the observer is not at rest anymore with respect to $B$.
Denoting the trace of the normal momentum surface density by $l$ which measures
the expansion of $B$ in time the boost-invariant QLE becomes
\begin{eqnarray}
E = \frac{1}{\kappa} \int_B d^2 x \sqrt{\sigma} \left ( \sqrt{k^2-l^2} - \sqrt{k_0^2-l_0^2} \right )
\end{eqnarray}
The most attractive feature of the boost-invariant QLE is certainly
its independence of the observer. It can attain complex values if
$B$ is located inside the event horizon of a black hole. Also, the
integrand of the unreferenced boost-invariant QLE is always positive (if real),
whereas the Brown-York QLE depends on the extrinsic curvature
of $B$ which can be positive or negative. 

Ultimately, the usefulness of the described quantities will depend on the
availability of applications. A more thorough review of the problems
associated with quasi-local energy can be found in \cite{lrr-2004-4}.

\section*{Acknowledgments}
Research support has been received from the National Science Foundation under
contract \#AST-0507813. The author also wishes to thank M.-T. Wang and S.-T.Yau
for discussions.

\bibliography{bib}

\begin{thebibliography}{10}

\bibitem{Bel:1958}
L~Bel.
\newblock {\em CR Acad. Sci. Paris}, \textbf{247}, 1094 {(1958)}.

\bibitem{Robinson:1958}
I.~Robinson.
\newblock {\em Kings College Lectures, unpublished}.
\newblock 1958.

\bibitem{Bel:1959}
L~Bel.
\newblock {\em CR Acad. Sci. Paris}, \textbf{248}, 1297 {(1959)}.

\bibitem{ADM:1962}
R.~Arnowitt, S.~Deser, and C.~W. Misner.
\newblock The dynamics of general relativity.
\newblock In L.~Witten, editor, {\em Gravitation: An Introduction to Current
  Research}, page 246, New York, 1962. Wiley.

\bibitem{Bondi1964}
A.~Trautman, F.~A.~E. Pirania, and H.~Bondi.
\newblock {\em Lectures on General Relativity}, volume~1.
\newblock Prentice-Hall, Englewood Cliffs, NJ, 1965.
\newblock Brandeis 1964 Summer Institute on Theoretical Physics.

\bibitem{Brown:1992br}
J.~David Brown and James~W. York, Jr.
\newblock Quasilocal energy and conserved charges derived from the
  gravitational action.
\newblock {\em Phys. Rev.}, \textbf{D47}, 1407--1419 {(1993)}.

\bibitem{Lundgren:2006fu}
Andrew~P. Lundgren, Bjoern~S. Schmekel, and James~W. York, Jr.
\newblock Self-renormalization of the classical quasilocal energy.
\newblock {\em Phys. Rev. D}, \textbf{75}, 084026 {(2007)}, gr-qc/0610088.

\bibitem{Blau:2007wj}
Matthias Blau and Blaise Rollier.
\newblock Brown-york energy and radial geodesics.
\newblock arXiv:0708.0321 [gr-qc].

\bibitem{Brown:1998bt}
J.~David Brown, S.~R. Lau, and Jr. York, James~W.
\newblock Canonical quasilocal energy and small spheres.
\newblock {\em Phys. Rev.}, \textbf{D59}, 064028 {(1999)}, gr-qc/9810003.

\bibitem{Schon:1981vd}
Richard Schon and Shing-Tung Yau.
\newblock Proof of the positive mass theorem. 2.
\newblock {\em Commun. Math. Phys.}, \textbf{79}, 231--260 {(1981)}.

\bibitem{Epp:2000zr}
Richard~J. Epp.
\newblock Angular momentum and an invariant quasilocal energy in general
  relativity.
\newblock {\em Phys. Rev. D}, \textbf{62}, 124018 {(2000)}, gr-qc/0003035.

\bibitem{Liu:2003bx}
Chiu-Chu~Melissa Liu and Shing-Tung Yau.
\newblock Positivity of quasilocal mass.
\newblock {\em Phys. Rev. Lett.}, \textbf{90}, 231102 {(2003)}, gr-qc/0303019.

\bibitem{lrr-2004-4}
Laszlo~B. Szabados.
\newblock Quasi-local energy-momentum and angular momentum in {G}{R}: A review
  article.
\newblock {\em Living Reviews in Relativity}, \textbf{7}.

\end{thebibliography}
\bibliographystyle{hunsrt}

\end{document}